\documentclass[12pt]{article}
 \usepackage{latexsym,epsfig,color,times}
\usepackage{colordvi}
\usepackage{amsmath}
\usepackage{latexsym,graphicx,color,times}
\usepackage[portrait,margin=1in]{geometry}

\newcommand{\be}{\begin{equation}}
\newcommand{\ee}{\end{equation}}

\def\hline{\centerline{\Red{\rule{6in}{1.0pt}}}}

\def\rfig#1{Fig.\ref{fig:#1}}

\def\req#1{(\ref{eq:#1})}

\def\macname{hankrs2}
\def\macname{hank_strauss}
\def\fignf12{./pix}
 
\def\figdirw{/Users/{\macname}/Documents/papers/disruption/rwtm/gnuplot}

\def\figdircc1{/Users/{\macname}/Documents/progs/m3dc1/plots}

\def\figdirw{./}

\begin{document}

\begin{center}
{\Large{\bf A Model of Tokamak Locked Mode  Disruptions}} \\
{\large\bf H. R. Strauss}
\end{center}
\begin{center}
%{{\bf H. Strauss}} \\ {HRS Fusion} \\
 {HRS Fusion} \\ {hank@hrsfusion.com} \\
%{{\bf  IVS-IPSTA 2021}}
\end{center}

\abstract{ \it
%There are several causes of tokamak disruptions. These contribute to 
%%disruption precursors, such as locked modes.
Locked modes are precursors to major disruptions.
During locked modes,
the temperature decreases in the plasma edge region.
This causes the current to contract. % and increases the internal inductance.
A model is given to analyze the MHD stability of contracted current equilibria.
If there is sufficient current contraction,  
resistive wall tearing modes are destabilized. This requires 
that the $q = 2$ surface be 
sufficiently close to the wall. The threshold conditions 
obtained in the model are
consistent with experimental observations of the conditions for
a thermal quench in a disruption.
}

\vspace{1cm}

Recent work has identified disruptions in JET \cite{jet21}, ITER \cite{iter21},
DIII-D \cite{d3d22}, and MST \cite{mst23} with resistive wall tearing modes (RWTMs)
\cite{finn95,gimblett,bondeson,betti}.
It was shown that RWTMs are able to cause a complete thermal quench.
An object of this paper is to show that the experimental  conditions for 
 tokamak locked mode
disruptions are also conditions for RWTM instability.  

Disruptions are generally preceded by precursors. 
This makes it possible to predict when disruptions occur.
Event chains \cite{sabbagh}
have been identified leading up to disruptions. 
Numerous causes of precursors in JET
have been identified \cite{devries11}, which lead to locked modes.
These  include
 neoclassical tearing modes (NTM) \cite{lahaye},
and  radiative cooling by impurities \cite{pucella}.
Locked modes are the main precursor of JET
disruptions, but they are not the instability causing the
thermal quench.  Rather,
the locked mode indicates an ``unhealthy" plasma which may disrupt
\cite{gerasimov2020}.
Locked modes are also disruption precursors in DIII-D \cite{sweeney2017,sweeney}.
The locked modes are tearing modes. They can overlap
and cause stochastic thermal transport in the plasma edge region.

During the locked mode phase, % prior to a major disruption,
edge transport and cooling modifies the edge temperature and current.
The drop in the edge temperature
causes the current to contract, while the total current stays 
constant. The result has been called \cite{schuller}  a
``deficient edge".  It has also been described \cite{sweeney}  as ``$T_{e,q2}$" collapse, a minor
disruption of the edge.
The contraction of the current is observed as an increase in the internal
inductance. A limiting internal inductance for disruptions has been observed in 
 JET \cite{wesson}, in TFTR \cite{cheng} and
in DIII-D \cite{sweeney2017}.

A condition for disruptions is that the $q = 2$ magnetic surface is sufficiently
near the plasma edge. This is been documented in DIII-D \cite{sweeney2017}.
It was found that disruptions require the $q = 2$ rational surface radius $r_s > 0.75 r_a,$ 
where $r_a$ is the plasma radius.

In the following, a  model is given to analyze the RWTM stability of contracted current equilibria.
It is shown that current contraction, and sufficiently large $r_s,$ are
conditions for RWTM
instability.

%The model equilibria are modified FRS \cite{frs} profiles.
%The magnetic field is
%\be b(r)   = (r/q_0) (1 + r^{2n})^{-1/n} \label{eq:b0} \ee
%where the plasma radius $r_a = 1,$ and $n$ is a  parameter that controls
%the current distribution. In FRS, a peaked profile has $n = 1,$
%%rounded, $n = 2,$ and flattened, $n = 4.$ In the model $n$ is a real
%number, not just an integer.
%Here $b = B_\theta / B_\phi$ in a large aspect ratio cylindrical cross section geometry.

%The current satisfies %\cite{frs,finn95,cheng}
%\be \frac{1}{x}\frac{d (x b) }{ d x} = j \label{eq:bj} \ee

The FRS current is
%\be j(x) = \frac{2}{q_0} \frac{1}{(1 + x^{2n})^{1 + 1/n}} \label{eq:jfrs} \ee
\be j(r) = \frac{2}{q_0} (1 + r^{2n})^{-(1 + 1/n)} \label{eq:jfrs} \ee
A peaked profile has $n = 1,$
rounded, $n = 2,$ and flattened, $n = 4.$ In this model $n$ is a real
number, not restricted to  an integer.
In order to cut off the current at $r = r_c,$ subtract a constant
$c_r$ with
\be c_r = (1 + r_c^{2n})^{-(1 + 1/n)} \label{eq:cr} \ee
where $r_c$ is the maximum radius of nonzero current.
\be
j(r)   =              \begin{cases}
 (2c_0/q_0)[ (1 + r^{2n})^{-(1 + 1/n)} - c_r] &  r < r_c  \\
  0  &  r \ge r_c.
\end{cases} \label{eq:cases} \ee
%\be j(x) = \frac{2c_0}{q_0} \left[\frac{1}{(1 + x^{2n})^{1 + 1/n}}
%- c_r \right]  \label{eq:jfrsc} \ee
%for $r \le r_c$
%\be c_r = \frac{1}{(1 + x_c^{2n})^{1 + 1/n}} \label{eq;cr} \ee
The factor  $c_0 = 1 / (1 - c_r)$ keeps $j(0)$ independent of $r_c.$
%\end{cases} \label{eq:cases-b} \ee
%and $q = r / b.$
%Modifying $b$ by subtracting $c_r r/q_0$ and multiplying by $c_0,$ gives $q = r / b,$
This gives a $q$ profile
\be
q(r)   =              \begin{cases}
 (q_0/c_0)[ (1 + r^{2n})^{-1/n} - c_r]^{-1}  &  r < r_c  \\
 q(r_c)(r/r_c)^2  &  r \ge r_c.
\end{cases} \label{eq:cases-q} \ee
Note that the total current is given by
\be I = r_a b(r_a) = r_a^2 / q_a = r_w^2 / q_w, 
 \label{eq:jtot} \ee
where $q = q_a$ at the plasma edge $r_a,$  or by $q_w,$ value at the wall $r_w.$

Sequences of equilibria during a precursor are modeled by keeping
$q_0 = 1,$ and by fixing $q_a$ to have  constant $I$.
During the sequence, $r_c$ is decreased. This causes the profile parameter  $n$ 
to increase,
in order to maintain constant $q_0, q_w.$ 
Current shrinking and broadening occur simultaneously.
The change in linear stability during this model sequence is investigated,
with both ideal and no wall boundary conditions. Resistive wall tearing modes,
%which  have been shown capable of causing a complete thermal quench, are
are tearing stable with an ideal wall, and unstable with no wall.

 The ideal wall tearing stability parameter $\Delta'_i$ and
the no wall tearing stability parameter $\Delta'_n$ are calculated in
cylindrical geometry. %The model is extended using a step current model
%\cite{frs,finn95,d3d22}.
RWTMs have  \cite{jet21,d3d22,mst23}
$\Delta'_i < 0$, and 
$\Delta'_{n} > 0.$

Linear magnetic perturbations satisfy \cite{finn95,cheng,frs,fkr}
\be \frac{1}{r}\frac{d}{dr}r\frac{d\psi}{dr} - \frac{m^2}{r^2} \psi  =
\frac{m}{r} \frac{dj}{dr} \frac{m/q - n}{[(m/q - n)^2 + m^2 \delta^2]} \psi
\label{eq:eq} \ee
where the singularity at the rational surface is regularized \cite{cheng},  with
$\delta = 10^{-4}.$
In case $r_c < r_s,$ the right side of \req{eq} vanishes for $r > r_c$, so there is
no singularity at $r_s$  and 
$\psi \propto r^{\pm 2}.$
Here $(m,n)$ are the poloidal and toroidal mode numbers of
a perturbation $\psi(r)\exp(im\theta - in \phi),$
using a large aspect ratio approximation.

Solving with a shooting method, there are two boundary conditions:
integrating outward from $r = 0,$ and inward from $r = r_w,$ the
wall radius.
The boundary conditions at the origin are
$\psi(0) = 0, d\psi/dr(0) = 0,$ since $\psi \sim r^m,$
with $m \ge 2.$
At the wall $r=r_w,$  an ideal wall boundary condition is
$\psi(r_w) = 0,$ $d\psi/dr(r_w) = 1.$
A resistive wall (or no wall)  boundary condition is
$\psi(r_w) = 1,$ $d\psi/dr(r_w) = -(m/r_w) \psi(r_w).$

The value of $\Delta'$ is calculated at $r_s$ at
which $q(r_s) = m/n,$
%\be \Delta' = \left(\frac{d \psi_+(r_s)}{dr} - \frac{d \psi_-(r_s)}{dr}\right)
%\frac{1}{\psi(r_s)}  \label{eq:deltap} \ee
\be \Delta' = \frac{\psi'_+(r_s) - \psi'_-(r_s)} {\psi(r_s)}  \label{eq:deltap} \ee
where $\psi' = d\psi/dr$,
$\psi_-$ is the solution integrated outward from $r = 0,$ and
$\psi_+$ is the solution integrated inward from $r = r_w.$
For an ideal wall, denote $\Delta' = \Delta_i,$ while for no wall,
$\Delta' = \Delta_n.$ The RWTM instability condition  is $\Delta_i \le 0,$
$\Delta_n \ge 0.$

\begin{figure}[h]
\vspace{.5cm}
\centering
 \includegraphics[height=5.0cm]{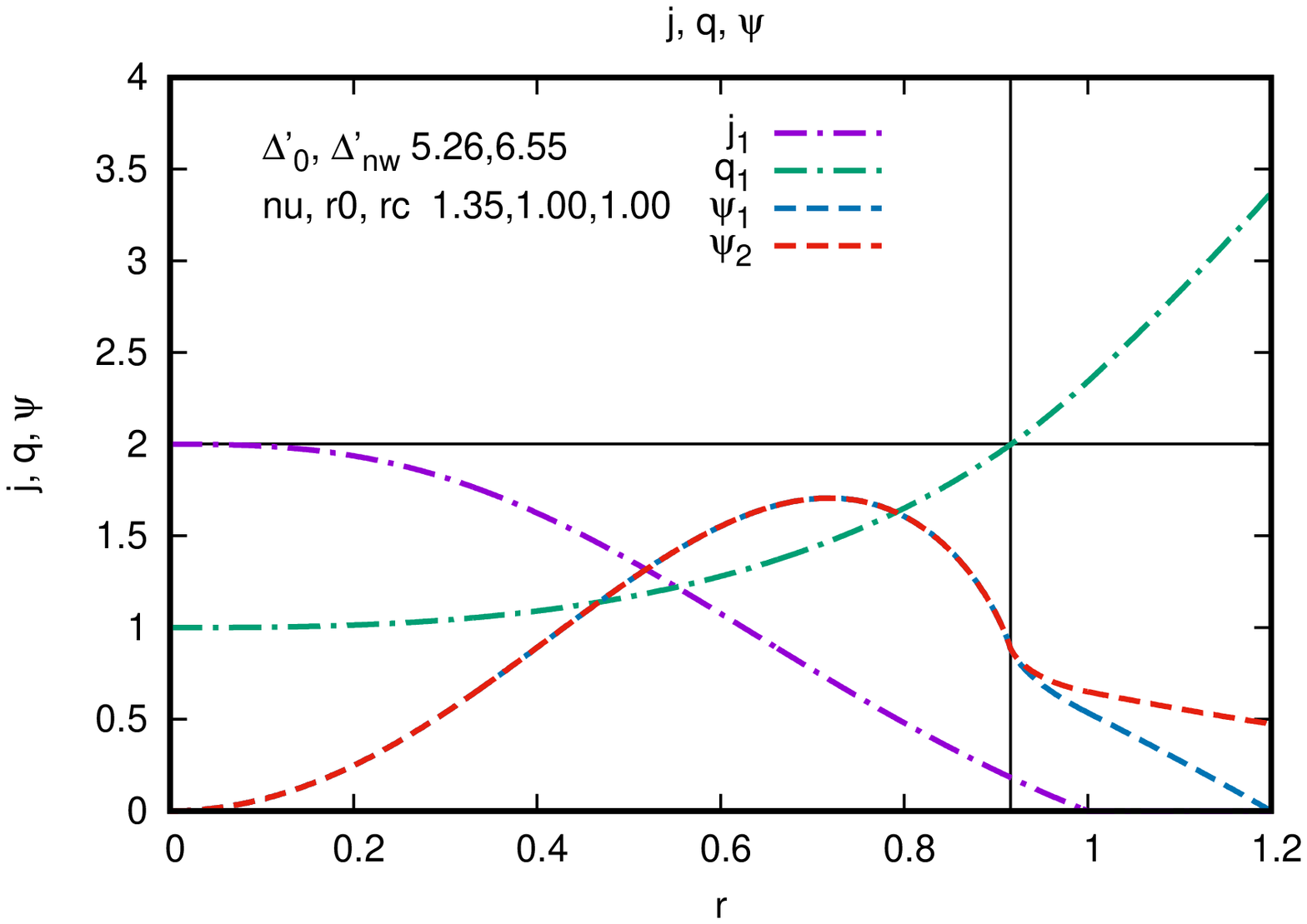}(a)
 \includegraphics[height=5.0cm]{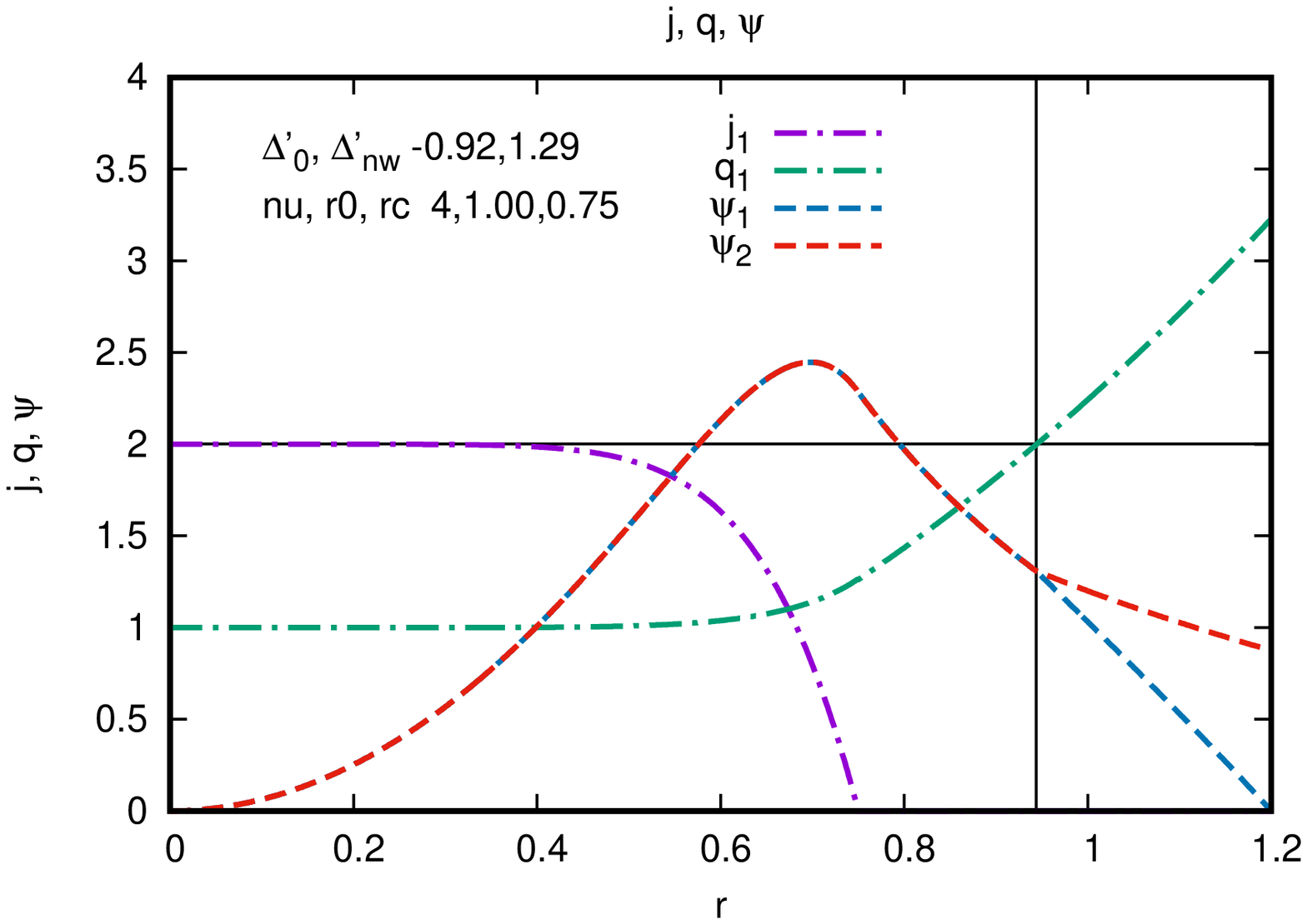}(b)
\caption{\it
$\psi,$ $j$, and $q,$  with $\psi$ for ideal $(\psi_1)$  and no wall
$(\psi_2)$. (a)
tearing mode unstable. The current is nonzero for $r < 1.$ (b) RWTM unstable.
The current
is non zero for $r  < r_c =  .75.$ The current profile is flattened
so the total current is almost the same as in (a). In both
cases $q_0 = 1.$
}
%\vspace{-3.0cm} % Adjust vertical figure spacing
 \label{fig:itlock}
 \end{figure}

The effect of the boundary conditions is illustrated in \rfig{itlock}(a),(b).
The plots show $j(r),$ $q(r)$ and $\psi(r)$ for
both ideal wall $(\psi_1)$ and resistive wall $(\psi_2)$.
The plasma boundary is $r_a = 1,$ and 
the wall is at $r_w = 1.2.$
The values of $\psi$ were normalized so that $\psi_+(r_s) = \psi_-(r_s).$
In each figure the two cases have the same profiles of $j$ and $q,$ as well
as the same $\psi_-$. The profiles of $\psi_+$ differ. The no wall
boundary condition produces a more positive value of $\Delta',$
\be \Delta'_{n} - \Delta'_i = \Delta'_x \ge 0. \label{eq:deltax1} \ee

\rfig{itlock}(a),(b) have different $j(r)$ profiles.
Both cases have approximately the same total current $J$
and have $q_0 = 1.$ It can be seen that $q(r_w)$ is approximately
the same. In \rfig{itlock}(a), $j$ is non zero for $r <  1.$
In \rfig{itlock}(b), $j$ is non zero for $r < r_c = 0.75.$
There is a marked difference in $\Delta'.$ 
The case
in \rfig{itlock}(a) is unstable to a tearing mode, while the second case
in \rfig{itlock}(b)
is unstable to a RWTM.
This supports the conjecture that suppressing the current in the plasma
edge region destabilizes the RWTM. The RWTM also requires that $r_s$ be
sufficiently close to $r_w,$ so that $\Delta'_i$ can become less than zero.

\begin{figure}[h]
\vspace{.5cm}
%\centering
\begin{center}
 \includegraphics[height=4.5cm]{\figdirw/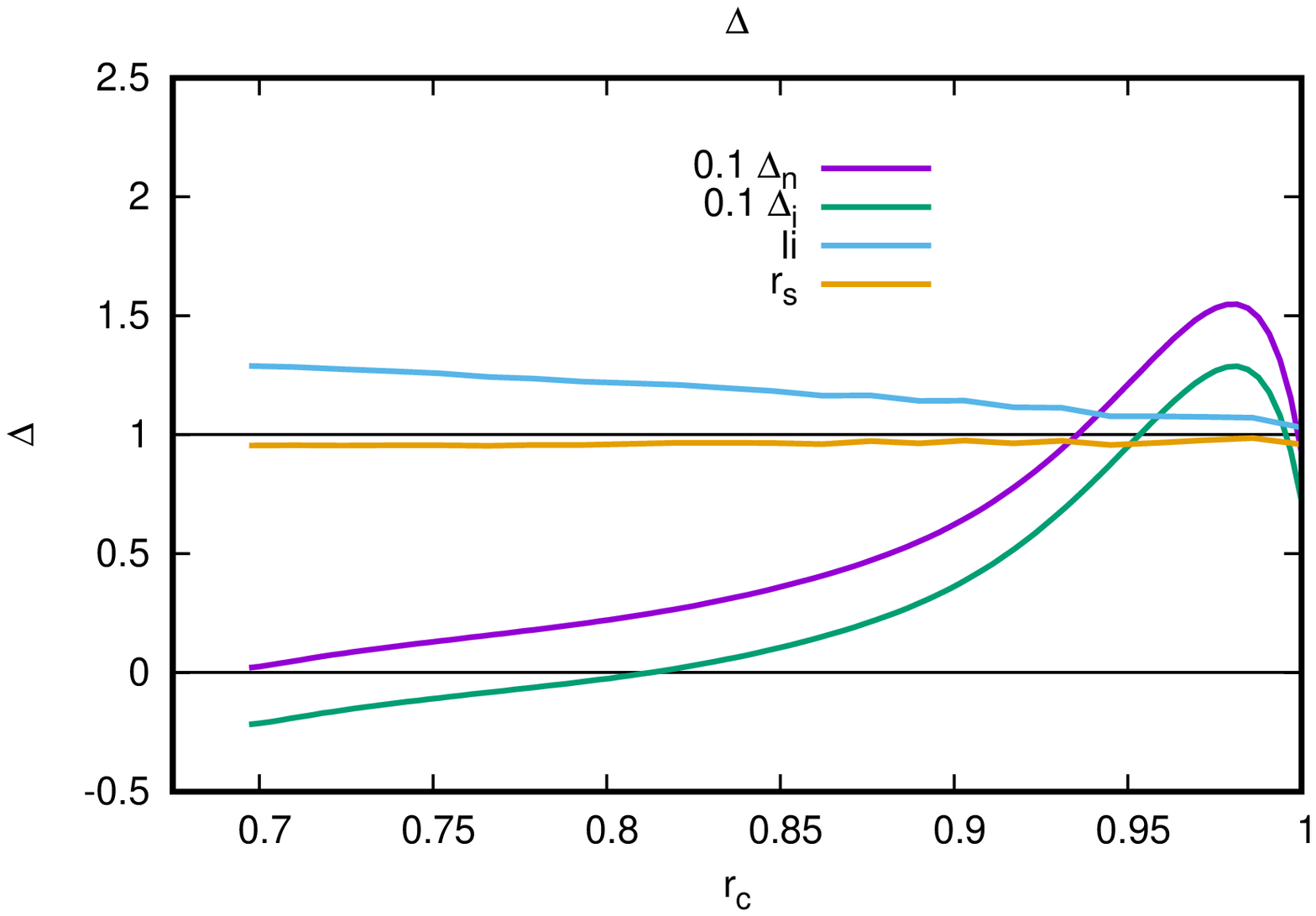}(a)
\includegraphics[height=4.5cm]{\figdirw/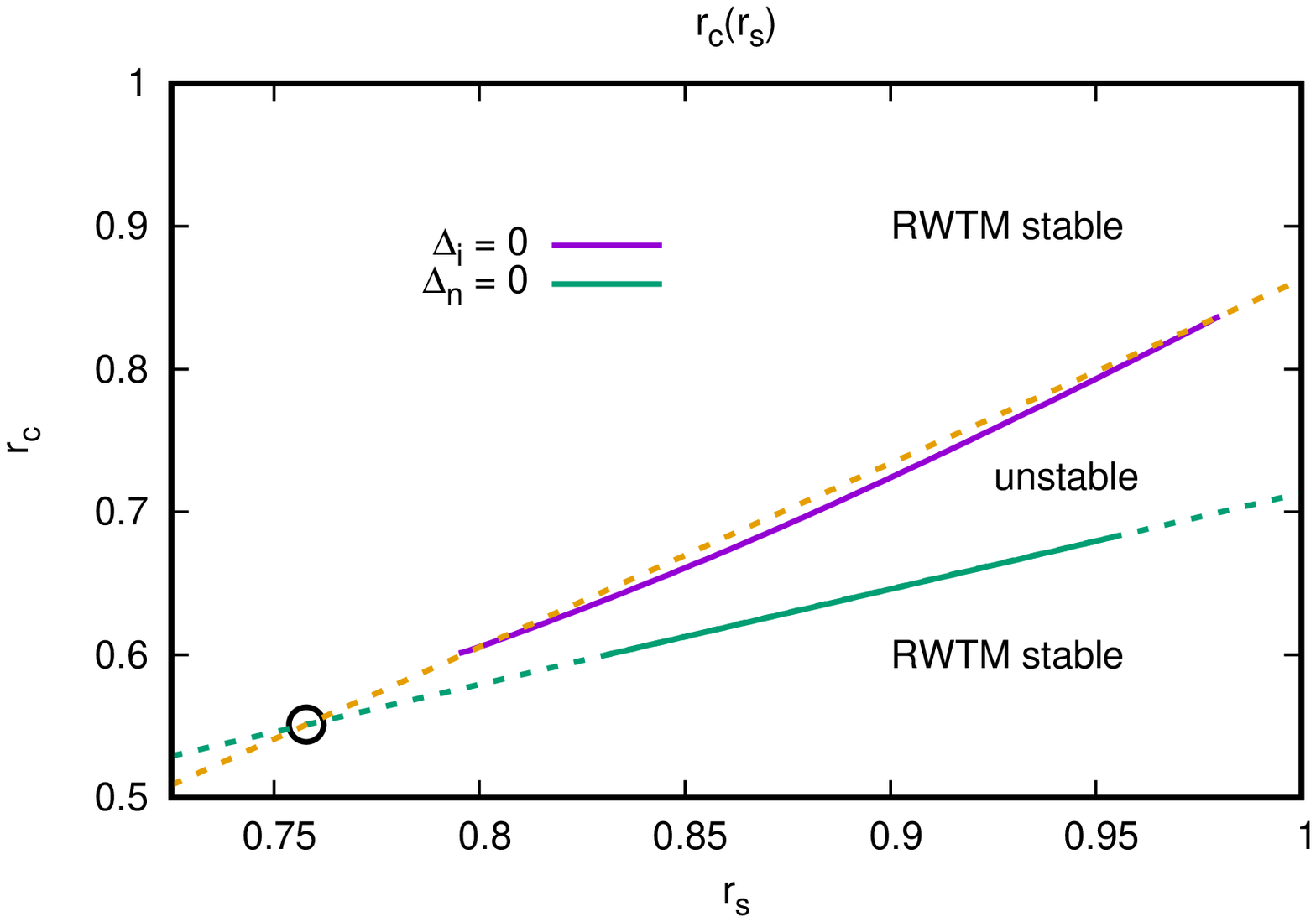}(b)
\end{center}
\caption{\it  
 (a)  $0.1 \Delta_n, 0.1 \Delta_i,$ $li,$ and $r_s$ as a function of
$r_c,$ for  $q_a = 2.2. $ $li$ increases as $r_c$ decreases.
$\Delta_i < 0$ for $r_c < 0.8,$ and $\Delta_n < 0$ for $r_c < 0.7.$
(b) $r_c$ %$li,$ 
as a function of
$r_s $ for which $\Delta_i \le 0, \Delta_n > 0 $,  and for which
 $\Delta_i \le 0,\Delta_n \le 0.$  
The $r_c$ curves are fitted with straight lines, which intersect at $r_s = 0.76.$
}
 \label{fig:qarc}
 \end{figure}

\rfig{qarc}(a) shows how $\Delta_i,\Delta_n$ vary with the current limiting radius
$r_c.$ The rational surface radius $r_s = .95$ is constant. 
As $r_c$ decreases, $li$ increases. The values of $\Delta_i,\Delta_n$ decrease, with
$\Delta_n > \Delta_i.$ Their values are multiplied by $0.1$ to fit in the plot.
For $r_c \le  0.8,$ $\Delta_i \le 0.$ This is the onset condition for a RWTM. 
For $r_c \le  0.7,$ $\Delta_n \le 0.$ This implies the RWTM is stabilized.
There is a range of $0.8 \ge r_c \ge 0.7$ in which the RWTM is unstable.

\rfig{qarc}(b)
shows how the marginal $\Delta_i,\Delta_n$ values vary with $r_s.$  
The critical values of
$r_c$ are found for both $\Delta_i =0, $ 
and for $\Delta_n =0. $
As in \rfig{qarc}(a) there is a
gap in $r_c$ between RWTM instability and stability. 
The $r_c$ curves are fit with straight lines, which intersect
at $r_s = 0.76.$ For $r_s < 0.76,$ RWTM is stable.
 This agrees well with a DIII-D database \cite{sweeney2017}.
%The conditions for RWTM instability in the model are consistent with
%the experimental conditions for disruption.
\begin{figure}[h]
\vspace{.5cm}
%\centering
\begin{center}
\includegraphics[height=4.5cm]{\figdirw/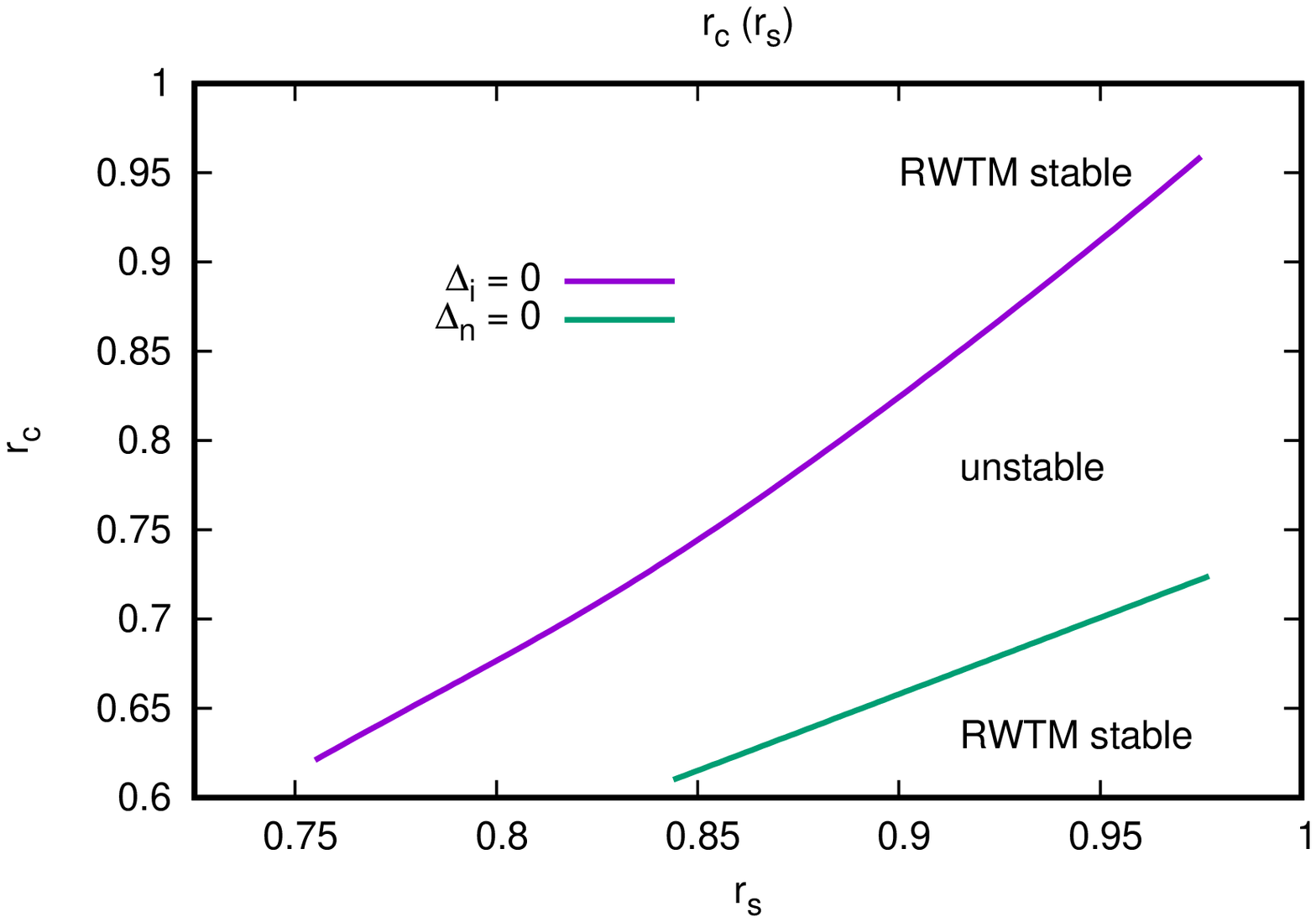}(a)
\includegraphics[height=4.5cm]{\figdirw/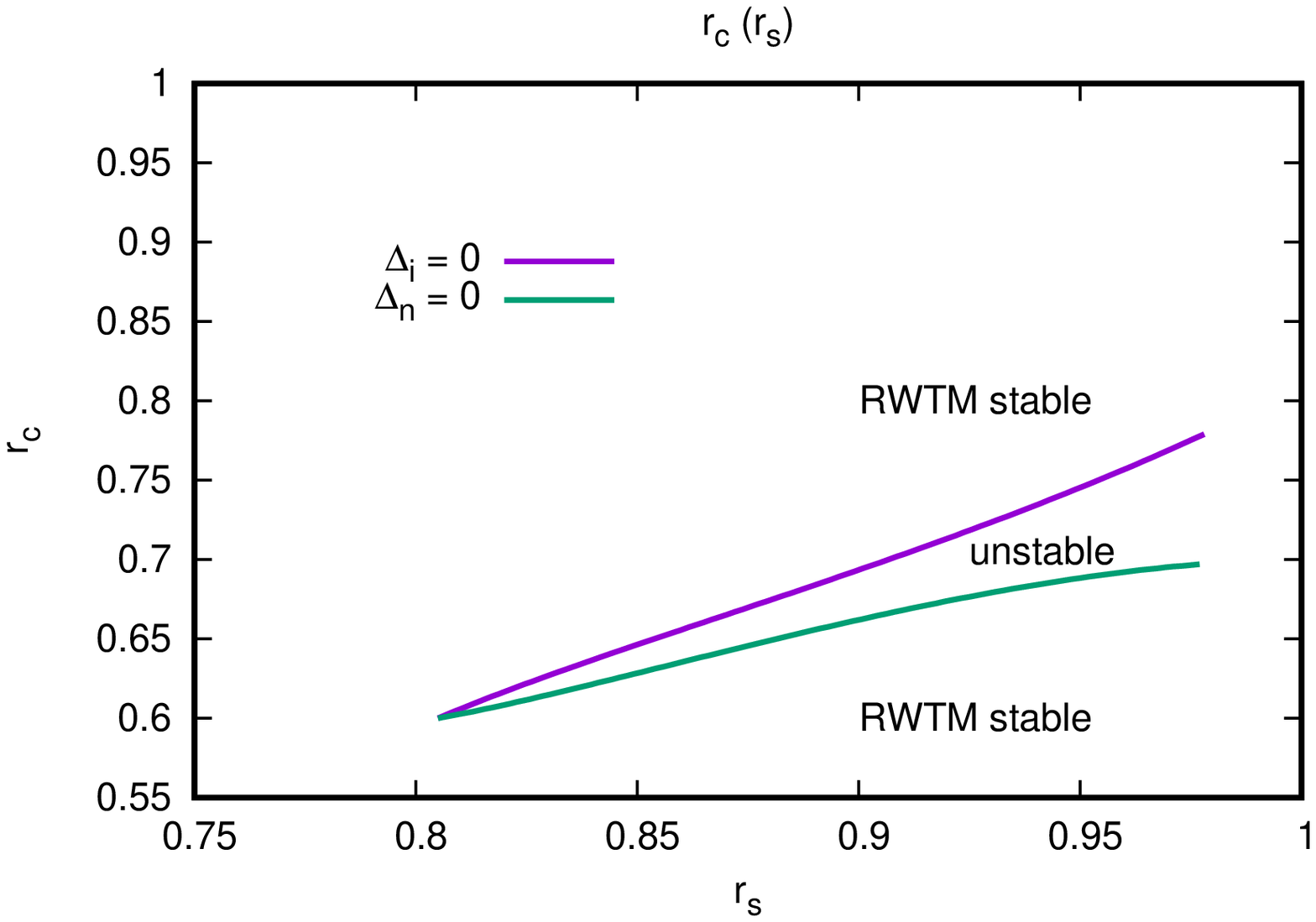}(b)
\end{center}
\caption{\it 
(a) $r_c$ %$li,$ 
as a function of
$r_s $ for which $\Delta_i \le 0, \Delta_n > 0, $ and for which
 $\Delta_i \le 0,\Delta_n \le 0.$ Here $r_w = 1.05,$
similar to MST. It is much more RWTM unstable than in \rfig{qarc}(b).
(b) the same, but with $r_w = 1.5.$ In this case there is less difference
between ideal and no wall boundary conditions, and the RWTM regime is small.
}
 \label{fig:qarc2}
 \end{figure}

When $r_s < 0.76,$ $\Delta' < 0$ for
both tearing and RWTMs. This implies a regime of stability.
It is possible that when $\Delta' \ll 0,$ kink modes or resistive kink modes 
are destabilized. Before that happens, the plasma must first evolve into the
region of RWTM instability, which could cause a disruption.

\rfig{qarc2} shows the effect of wall location in the model. Intuitively,
the closer the wall is to the plasma, the larger is the 
RWTM regime. The further away the wall is located, the difference between
ideal and no wall boundary conditions is smaller. \rfig{qarc2}(a) is similar
to \rfig{qarc}(b), with $r_w = 1.05.$ The RWTM unstable regime is enlarged.
This is consistent with MST, which should be quite unstable to RWTMs. The
case $r_w = 1.2$ is comparable to DIII-D, in which $r_s > 0.75$ for disruptions.
\rfig{qarc2} (b) shows the case $r_w = 1.5.$ The RWTM regime is small. 
Comparing \rfig{qarc}(b), \rfig{qarc2}(b), the minimum $r_s$ for RWTM instability
increases as $r_w$ increases, which is intuitively reasonable.

Although the model is relatively simple, it gives results qualitatively
and even quantitively consistent with experiment. One possible improvement would
be to include some current outside the main current channel. This would be more
realistic and would lower the value of $li$. However, it adds an extra
parameter which would make the model unduly complicated.

To summarize,
disruption precursors  have many causes, leading to locked modes in ITER and DIII-D.
During precursors, the edge temperature is  reduced, causing the current to  contract.
This is observed as an increase of internal inductance.
Experimentally, disruptions have onset when internal inductance  is greater than a threshold.
Disruption onset also requires the $q = 2$ rational surface to be  greater than a critical value. 
These onset conditions are consistent with RWTM destabilization. 
A model set of equilibria  is given which includes  current contraction,
while maintaining constant total current and $q = 1$ on axis.
%as well as steepening of the cuurent gradient.
Linear MHD equations are solved with ideal wall and no wall boundary
conditions. No wall boundary conditions always make the tearing mode more unstable
than ideal wall boundary conditions.
If a tearing mode is stable with and ideal wall and unstable with no wall,
it is a resistive wall tearing mode. %which has been shown capable of causing a
%complete thermal quench in a disruption. 
%The model is consistent with experimental  thresholds and with  RWTM onset.
For a sufficiently large $q = 2$ radius, which depends on the wall radius,
shrinking the current radius $r_c$ destabilizes the RWTM.
Further shrinking of  $r_c$  stabilizes the RWTM, which exists in a
range of $r_c$ values. Even further shrinking of $r_c$  might destabilize
kink modes, but this is outside the scope of the model.   

%Future work will study precursor experimental reconstructions. 

%\begin{figure}[h]
%\vspace{.5cm}
%\begin{center}
% \includegraphics[height=6.5cm]{\figdirsh/sweeneyfig21-2017.eps}
%\end{center}
%\caption{\it  
%Disruption occurrence depends on
%normalized $q=2$ radius,
%$\rho_{q2} > .75$ %in DIII-D
%This is  the onset condition in a database of DIII-D disruptions
%$\rho_{q2} = (r_{w}/a ) x_s \approx 1.2 x_s.$
%The stability boundary is $x_s \approx 0.625.$  
%From \cite{sweeney2017}. 
%}
% \label{fig:swe}
% \end{figure}

{\bf Acknowledgement} The support of USDOE grant DE-SC0020127 is acknowledged.


\begin{thebibliography}{9}
\bibitem{jet21}
H. Strauss and JET Contributors,
Effect of Resistive Wall on Thermal Quench in JET Disruptions,
Phys. Plasmas \textbf{28}, 032501 (2021) %; doi: 10.1063/5.0038592
\bibitem{iter21} H. Strauss, Thermal quench in ITER disruptions,
 Phys. Plasmas \textbf{28} 072507 (2021) %; https://doi.org/10.1063/5.0052795}
\bibitem{d3d22} H. Strauss, B. C. Lyons, M. Knolker,
Locked mode disruptions in DIII-D and application to ITER,
Phys.  Plasmas \textbf{29}  112508 (2022). % doi: 10.1063/5.0107048
\bibitem{mst23} H. R. Strauss,  B. E. Chapman, N. C. Hurst,
MST Resistive Wall Tearing Mode Simulations, %arXiv:2302.11926  (2023)
http://arxiv.org/abs/2302.11926 (2023).
\bibitem{finn95} John A. Finn,
Stabilization of ideal plasma resistive wall modes in cylindrical
geometry: the effect of resistive layers, 
Phys. Plasmas 2, 3782 (1995)
\bibitem{gimblett} C.G. Gimblett,
On free boundary instabilities induced by a resistive wall,
Nucl. Fusion  \textbf{26} 617 (1986)
\bibitem{bondeson} A. Bondeson and M. Persson,
Stabilization by resistive walls and q-limit disruptions in tokamaks,
Nucl. Fusion \textbf{28} 1887 (1988)
\bibitem{betti} R. Betti,
Beta limits for the n = 1 mode in rotating - toroidal - resistive plasmas
surrounded by a resistive wall, Phys. Plasmas \textbf{5} 3615 (1998).
\bibitem{sabbagh}
A. Sabbagh,   J. W. Berkery,   Y. S. Park,   J. Butt, J. D. Riquezes,   J. G. Bak,   R. E. Bell,  
 L. Delgado-Aparicio,
S. P. Gerhardt, C. J. Ham,   J. Hollocombe,   J. W. Lee, J. Kim,   A. Kirk,   J. Ko, W. H. Ko, 
L. Kogan, J. H. Lee,   A. Thornton, and S. W. Yoon,
Disruption event characterization and forecasting in tokamaks, 
Physics of Plasmas \textbf{30}, 032506 (2023)
\bibitem{devries11}
P.C. de Vries, M.F. Johnson, B. Alper, P. Buratti,
T.C. Hender, H.R. Koslowski, V. Riccardo and JET-EFDA Contributors,
Survey of disruption causes at JET, Nucl. Fusion \textbf{51}   053018  (2011).`
\bibitem{lahaye} 
R.J. La Haye, C. Chrystal , E.J. Strait , J.D. Callen, C.C. Hegna, E.C. Howell , M. Okabayashi
and R.S. Wilcox,
Disruptive neoclassical tearing mode seeding in DIII-D with implications for ITER,
 Nucl. Fusion 62 056017 (2022).
\bibitem{pucella} G. Pucella, P. Buratti, E. Giovannozzi, E. Alessi,
F. Auriemma, D. Brunetti, D. R. Ferreira, M. Baruzzo,
D. Frigione, L. Garzotti, E. Joffrin, E. Lerche, P. J. Lomas, S. Nowak, L. Piron,
F. Rimini, C. Sozzi, D. Van Eester, and JET Contributors,
Tearing modes in plasma termination on JET:
the role of temperature hollowing and edge cooling,
 Nucl. Fusion \textbf{61} 046020 (2021)
%\bibitem{devries16}
%P.C. de Vries, G. Pautasso, E. Nardon, P. Cahyna, S. Gerasimov, 
%J. Havlicek, T.C. Hender, G.T.A. Huijsmans, M. Lehnen, M. Maraschek, T. Markovic,
% J.A. Snipes and the COMPASS Team, the ASDEX Upgrade Team and JET Contributors,
%Scaling of the MHD perturbation amplitude required to trigger a disruption
%and predictions for ITER, Nucl. Fusion \textbf{56}  026007 (2016)
%\bibitem{gerasimov19} 
%S.N. Gerasimov, P. Abreu, G. Artaserse, P. Buratti, I.S. Carvalho, E. de la Luna, 
%A.R. Field, E. Giovannozzi, T.C. Hender, R.B. Henriques, P.J. Lomas, E. Matveeva, S. Moradi,
%L. Piron, F.G. Rimini, H. Sun, G. Szepesi, L.E. Zakharov, and JET Contributors,
%Locked mode and disruptions in JET-ILW,
%46th EPS Conference on Plasma Physics, P1.1056 (2019).
%\bibitem{bandyopadhyay} 
%Indranil Bandyopadhyay, Matteo Barbarino, Amitava Bhattacharjee, Nicholas Eidietis , Alexander Huber ,
% Akihiko Isayama, Jayhyun Kim, Sergey Konovalov, Michael Lehnen, Eric Nardon,
%Gabriella Pautasso, Cristina Rea, Carlo Sozzi, Fabio Villone and Long Zeng,
%Summary of the IAEA technical meeting on plasma disruptions and their mitigation,
%Nucl. Fusion \textbf{61}  077001 (2021).
\bibitem{gerasimov2020}
S.N. Gerasimov, P. Abreu, G Artaserse, M. Baruzzo, P. Buratti, I.S. Carvalho,
I.H. Coffey, E. de la Luna, T.C. Hender, R.B. Henriques, R. Felton, S. Jachmich,
U. Kruezi, P.J. Lomas, P. McCullen, M. Maslov, E. Matveeva, S. Moradi, L. Piron,
F.G. Rimini, W. Schippers, G. Szepesi, M. Tsalas, L.E. Zakharov and JET Contributors,
Overview of disruptions with JET-ILW,
Nucl. Fusion \textbf{60}  066028 (2020).
\bibitem{sweeney2017} R. Sweeney, W. Choi, R. J. La Haye, S. Mao, K. E. J.
Olofsson, F. A. Volpe, and the DIII-D Team,
Statistical analysis of m/n = 2/1 locked and quasi - stationary modes
with rotating precursors in DIII-D,
Nucl. Fusion 57 0160192 (2017) 
\bibitem{sweeney} R. Sweeney, W. Choi, M. Austin, M. Brookman, V. Izzo, M. Knolker,
R.J. La Haye, A. Leonard , E. Strait, F.A. Volpe and The DIII-D Team,
Relationship between locked modes and thermal quenches in DIII-D,
Nucl. Fusion 58 056022 (2018). %; doi:10.1088/1741-4326/aaaf0a
\bibitem{schuller}
F.C. Schuller, Disruptions in tokamaks, Plasma Phys. Controlled Fusion \textbf{37}, A135 (1995).
\bibitem{wesson}
J.A. Wesson, R.D. Gill, M. Hugon, F.C. Schuller, J.A. Snipes, DJ. Ward, D.V. Bartlett,
D.J. Campbell, P.A. Duperrex, A.W. Edwards, R.S. Granetz, N.A.O. Gottardi, T.C. Hender,
E. Lazzaro, P.J. Lomas, N. Lopes Cardozo, K. F. Mast,
M.F.F. Nave, N.A. Salmon, P. Smeulders, P.R. Thomas, B.J.D. Tubbing, M.F. Turner, A. Weller,
Disruptions in JET, Nucl. Fusion \textbf{29} 641 (1989).
%\bibitem{garofalo} 
%A.M. Garofalo, G.L. Jackson, R.J. La Haye, M. Okabayashi3, H. Reimerdes, E.J. Strait,
%J.R. Ferron, R.J. Groebner, Y. In, M.J. Lanctot, G. Matsunaga, G.A. Navratil, W.M. Solomon,
%H. Takahashi3, M. Takechi, A.D. Turnbull and the DIII-D Team,
%Stability and control of resistive wall modes in high beta, low rotation DIII-D plasmas,
%Nucl. Fusion \textbf{47} 1121–1130 (2007).
%\bibitem{izzo}
% V. A. Izzo, 
%A numerical investigation of the effects of impurity penetration depth on disruption 
%mitigation by massive high-pressure gas jet,
%Nucl. Fusion \textbf{46} 541 (2006). %doi:10.1088/0029-5515/46/5/006
% V. A. Izzo, D. G. Whyte, R. S. Granetz, P. B. Parks,
%E. M. Hollmann, L. L. Lao, J. C. Wesley,
%Magnetohydrodynamic simulations
%of massive gas injection int Alcator C - Mod and DIII-D plasmas,
%Phys. Plasmas \textbf{15}, 056109 (2008).
%\bibitem{lehnen}
%M.Lehnen, K.Aleynikova, P.B.Aleynikov D.J.Campbell, P.Drewelow, N.W.Eidietis,
%Yu.Gasparyan, R.S.Granetz, Y.Gribov, N.Hartmann, E.M.Hollmann,  V.A.Izzo, S.Jachmich,
%S.-H.Kim, M.Kočan, H.R.Koslowski, D.Kovalenko, U.Kruezi, A.Loarte, S.Maruyama,
%G.F.Matthews, P.B.Parks, G.Pautasso, R.A.Pitts, C.Reux, V.Riccardo, R.Roccella,
%J.A.Snipes, A.J.Thornton, P.C.de Vries, EFDA JET contributors,
%Disruptions in ITER and strategies for their control and mitigation,
%Journal of Nuclear Materials, \textbf{463}, 39 (2015)
%\bibitem{gates} D. A. Gates, D. P. Brennan, L. Delgado-Aparicio, and R. B. White,
%The tokamak density limit: A thermo-resistive disruption mechanism,
%Phys. of Plasmas 22, 060701 (2015). % ; doi: 10.1063/1.4922472
%\bibitem{teng}
%Q. Teng, N. Ferraro, D.A. Gates, R.B. White, Nonlinear simulations 
%of thermo-resistive tearing mode formalism of the density limit, Nucl. Fusion (2018).
%\bibitem{mst} N. C. Hurst, B. E. Chapman, A. F. Almagri, B. S. Cornille, S. Z. Kubala,
% K. J. McCollam, J. S. Sarff, C. R. Sovinec, J. K. Anderson, D. J. Den Hartog, C. B. Forest,
% M. D. Pandya, and W. S. Solsrud,
%Self-organized magnetic equilibria in tokamak plasmas with very low edge safety factor,
%Phys. Plasmas  \textbf{29} 080704 2022.
%\bibitem{iter1999}
%%ITER Physics Expert Group on Disruptions, Plasma Control, and MHD: 
%S. Mirnov, J. Wesley,
% N. Fujisawa, Yu. Gribov, O. Gruber, T. Hender, N. Ivanov, S. Jardin, J. Lister,
%F. W. Perkins, M. Rosenbluth, N. Sauthoff, T. Taylor, S. Tokuda, K. Yamazaki, R. Yoshino,
%A. Bondeson, J. Conner, E. Fredrickson, D. Gates, R. Granetz, R. La Haye, J. Neuhauser,
%F. Porcelli, 
%D.E. Post, N.A. Uckan, M. Azumi, D.J. Campbell, M. Wakatani, W.M. Nevins, M. Shimada, 
%J. Van Dam,
%ITER Physics Basis
%Chapter 3: MHD stability, operational limits and disruptions,
%Nuclear Fusion, \textbf{39},  2251 (1999)
\bibitem{cheng}
C. Z. Cheng,  P. Furth and A. H. Boozer,
MHD stable regime of the Tokamak,
Plasma Phys. Control. Fusion 29 351 (1987).
\bibitem{frs} H. P. Furth, P. H. Rutherford, and H. Selberg,
 Tearing mode in the cylindrical tokamak,
Physics of Fluids \textbf{16} 1054 (1973
\bibitem{fkr} H. P. Furth, J. Killeen, and M. N. Rosenbluth,
Finite-Resistivity Instabilities of a Sheet Pinch,
Phys. Fl. \textbf{6}, 459 (1963).


\end{thebibliography}
\end{document}